# Graph Neural Networks for User Satisfaction Classification in Human-Computer Interaction

Rui Liu
University of Melbourne
Melbourne, Australia

Runsheng Zhang
University of Southern California
Los Angeles, USA

Shixiao Wang*
School of Visual Arts
New York, USA

*Abstract-This study focuses on the problem of user satisfaction classification and proposes a framework based on graph neural networks to address the limitations of traditional methods in handling complex interaction relationships and multidimensional features. User behaviors, interface elements, and their potential connections are abstracted into a graph structure, and joint modeling of nodes and edges is used to capture semantics and dependencies in the interaction process. Graph convolution and attention mechanisms are introduced to fuse local features and global context, and global pooling with a classification layer is applied to achieve automated satisfaction classification. The method extracts deep patterns from structured data and improves adaptability and robustness in multi-source heterogeneous and dynamic environments. To verify effectiveness, a public user satisfaction survey dataset from Kaggle is used, and results are compared with multiple baseline models across several performance metrics. Experiments show that the method outperforms existing approaches in accuracy, F1-Score, AUC, and Precision, demonstrating the advantage of graph-based modeling in satisfaction prediction tasks. The study not only enriches the theoretical framework of user modeling but also highlights its practical value in optimizing human-computer interaction experience.*

*Keywords: User satisfaction judgment, graph neural network, attention mechanism, human-computer interaction*

## I. Introduction

In the context of deep integration between digitalization and intelligence, human-computer interaction has shifted from simple information transmission to intelligent interaction centered on user experience and satisfaction. With the rapid development of multimodal technologies, natural language processing, and immersive interaction methods, user expectations of interactive systems are no longer limited to functional implementation[1]. They now extend to emotional resonance, cognitive load, and overall satisfaction during use. User satisfaction is not only an important indicator of the success of interaction design but also a key factor in determining whether a system can be accepted and used continuously. How to accurately model and evaluate user satisfaction has become a core research issue in the field of human-computer interaction, carrying both theoretical and practical importance[2].

In complex and dynamic interaction environments, user satisfaction is shaped by multiple factors. These factors include individual differences and user goals, as well as task completion efficiency, interface usability, system response speed, and the coherence and predictability of interaction. Traditional methods, such as questionnaires, interviews, and observations, can capture subjective perceptions, but they face difficulties in handling large-scale, multi-source, and dynamic data. They are also unable to reflect real-time changes in the interaction experience. Therefore, how to apply intelligent modeling methods to describe user satisfaction in an objective, dynamic, and fine-grained manner has become an urgent challenge[3].

The emergence of graph neural networks offers a new perspective for modeling user satisfaction. Compared with traditional methods, graph neural networks are better suited for handling complex relationships and multi-level dependencies between users and systems [4-6]. They represent interactions through nodes and edges, capturing hidden interaction patterns. In human-computer interaction scenarios, user behavior sequences, associations between interface elements, and even indirect interactions among multiple users can be abstracted into graph structures [7-9]. By propagating and aggregating information in these structures, graph neural networks can characterize individual behaviors while revealing implicit contextual semantics and global dependencies, thus providing a more precise foundation for satisfaction evaluation[10].

Introducing graph neural networks into user satisfaction assessment has methodological innovation and practical value. On the one hand, this approach overcomes the limitations of linear and sequential models, maintaining strong robustness and adaptability in complex environments. On the other hand, it supports the integration of heterogeneous data sources, such as click behavior, dwell time, voice commands, and visual attention, to reflect the multidimensional attributes of user experience. This cross-modal and multi-level modeling provides a solid technical basis for adaptive interaction systems, enabling them to perceive user states and needs more proactively and intelligently.

In summary, research on user satisfaction evaluation based on graph neural networks represents an important direction in human-computer interaction and a critical step toward improving intelligent interaction experience. From an academic perspective, it enriches the theoretical framework of user modeling and deepens the understanding of satisfaction formation mechanisms. From an application perspective, it provides reliable support for scenarios such as personalized recommendation, intelligent customer service, educational platforms, and virtual reality. As human-computer interaction continues to evolve toward intelligence and emotional awareness, building a satisfaction evaluation framework oriented toward user experience will become an important path for iterative upgrades and sustainable development of interactive systems.

## II. RELATED WORK

In studies of user satisfaction evaluation, early approaches mainly relied on statistical analysis and questionnaire-based methods. These approaches collected subjective feedback from users after system use and built simple association models between satisfaction and interaction behavior[11]. Although they could reveal user preferences to some extent, they suffered from strong subjectivity, limited data scale, and poor ability to capture dynamic changes in interaction. With the increasing complexity of human-computer interaction systems, single questionnaires or behavioral indicators can no longer provide a full picture of user experience in different contexts. This limitation has driven academic research toward more automated and objective methods.

With the rapid development of machine learning, methods based on feature extraction and classification models have gradually become mainstream [12]. These methods usually relied on hand-crafted features extracted from interaction logs, operation traces [13], or dwell time, and then used traditional classifiers to evaluate satisfaction. Such approaches improved objectivity and automation, and partially addressed the shortage of subjective data. However, hand-crafted features often failed to capture the complexity of user interactions. In particular, when multimodal data and nonlinear interaction patterns are prevalent, their effectiveness was constrained by feature selection and modeling capacity[14].

In recent years, the introduction of deep learning has significantly expanded the scope of user satisfaction research. By using convolutional neural networks, recurrent neural networks, and attention-based models, researchers could automatically learn high-level feature representations directly from raw data [15], reducing reliance on feature engineering. These methods captured complex temporal and contextual information and showed strong adaptability in processing multimodal data [16]. In particular, attention mechanisms enabled models to identify key factors in long interaction sequences [17], improving the accuracy and interpretability of satisfaction prediction. However, such models often assumed inputs in sequence or matrix form, which limited their ability to capture potential non-Euclidean relationships between users and systems[18].

Against this background, graph neural networks have emerged as a promising direction for user satisfaction evaluation. By modeling interaction objects, behavioral sequences, and their associations as graph structures, graph neural networks integrate local features and global context through information propagation across nodes and edges [19]. Compared with traditional methods [20], they provide a natural way to represent complex relational patterns, such as hierarchical dependencies between interface elements, latent connections among user behaviors [21], and interactive effects among multiple users. This structured representation improves the accuracy of satisfaction evaluation and offers new perspectives on the mechanisms of user experience formation. With their success in recommendation systems, sentiment analysis, and interaction modeling, graph neural networks show strong potential in user satisfaction research and lay a solid foundation for advancing human-computer interaction experience.

## III. METHOD

In this study, the user satisfaction assessment task is formalized as a node classification problem on a graph structure. First, assume that the interaction process consists of a series of user behaviors and interface elements, which can be abstracted as a directed graph $G = (V, E)$, where V represents the node set, including user states and interaction elements, and E represents the edge set, which characterizes the semantic or behavioral dependencies between nodes. Each node $v_i \in V$ has an initial feature vector $x_i \in R^d$. To obtain a high-order representation of the node, a graph convolution operation is used to propagate and aggregate the node features. The update formula is:

$$h_i^{(l+1)} = \sigma( \sum_{j \in N(i)} \frac{1}{c_{ij}} W^{(l)} h_j^{(l)} ) \qquad (1)$$

Where $h_i^{(l)}$ is the node representation of the lth layer, $W^{(l)}$ is the trainable parameter, $N(i)$ is the neighbor set of node i, $c_{ij}$ is the normalization constant, and $\sigma(\cdot)$ is the nonlinear activation function. The model architecture is shown in Figure 1.

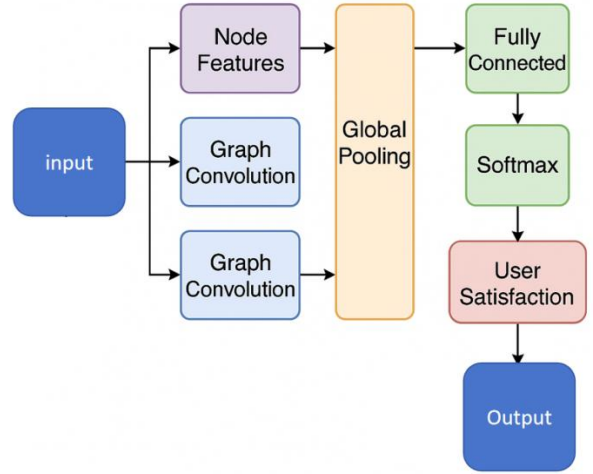

Figure 1. Overall model architecture

Based on graph convolution, the attention mechanism is introduced to improve the ability to model key interaction patterns. Specifically, the importance of information dissemination is dynamically adjusted by calculating the attention weights between nodes. For any edge $(i, j)$, the attention coefficient can be expressed as:

$$a_{ij} = \frac{\exp(LeakyRELU(a^T[Wh_i \| Wh_j]))}{\sum_{k \in N(i)} \exp(LeakyRELU(a^T[Wh_i \| Wh_j]))} \quad (2)$$

Where a is a learnable parameter vector, and ∥ represents a vector concatenation operation. Through this mechanism, the model can highlight interactive features that have a stronger impact on user satisfaction and weaken noise signals.

To further capture global dependencies, this study introduces a global pooling operation into the high-level structure of the graph representation to generate a discriminative overall vector representation [22]. Let $h_i^{(L)}$ be the node representation after multi-layer propagation and attention aggregation, then the global embedding of the entire graph can be expressed as:

$$z_G = READOUT(\{h_i^{(L)} \mid i \in V\}) \quad (3)$$

Where $READOUT(\cdot)$ can be an average pooling, maximum pooling, or attention pooling function. The vector $z_G$ combines the multi-layer features of the user behavior sequence and the interactive elements, and can effectively represent the user's overall experience during the interaction process.

In terms of training objectives, the satisfaction discrimination task is modeled as a binary classification problem, distinguishing between "satisfactory" and "unsatisfactory" interaction experiences. The model's final prediction is achieved through a fully connected layer and a softmax function, specifically:

$$\hat{y} = \text{Softmax}(W_o z_G + b_o) \quad (4)$$

Where $W_o$ and $b_o$ are the parameters of the output layer, and $\hat{y} \in R^2$ represents the predicted category distribution. To optimize the model, the cross-entropy loss function is introduced, which is defined as:

$$L = -\sum_{c=1}^{C} y_c \log \hat{y}_c \quad (5)$$

Where $y_c$ is the one-hot encoding of the true label, and $\hat{y}_c$ is the model's predicted probability. This objective function can help the model learn discriminative features that distinguish different satisfaction categories, thereby improving overall prediction performance.

IV. EXPERIMENTAL RESULTS

*A. Dataset*

The dataset used in this study comes from the Kaggle platform and belongs to a publicly available user satisfaction survey. It contains a large number of user feedback records across different services and interaction scenarios. The dataset includes questionnaire evaluations, rating levels, and multidimensional features related to interaction experience. The data types include numerical indicators such as rating scales and behavioral statistics, as well as categorical information such as usage scenarios and interaction channels. These features provide rich input for satisfaction modeling.

In terms of scale, the dataset comprises tens of thousands of user feedback records, ensuring the reliability of statistical analysis and model training. Its extensive coverage encompasses diverse experiences from various user groups and varied behaviors across multiple interaction scenarios. This comprehensive representation allows the dataset to accurately reflect the formation mechanism of user satisfaction. The large scale and multidimensional nature of the dataset make it not only suitable for classification tasks but also for more in-depth pattern mining of user satisfaction.

In addition, the dataset has been anonymized. Sensitive information has been removed, and only behavioral and feedback features meaningful for modeling have been retained. This ensures compliance and provides a clean and reusable data source for subsequent experiments. With its openness and accessibility, the dataset has become an important resource for satisfaction prediction and human-computer interaction analysis. It also offers a unified basis for comparison and validation of different methods.

*B. Experimental Results*

This paper also gives the comparative experimental results, as shown in Table 1.

Table1. Comparative experimental results

| Model | ACC | F1-Score | AUC | Precision |
|---|---|---|---|---|
| GCN[23] | 0.842 | 0.835 | 0.873 | 0.828 |
| GAT[24] | 0.857 | 0.846 | 0.884 | 0.841 |
| Transformer[25] | 0.865 | 0.852 | 0.892 | 0.849 |
| 1DCNN[26] | 0.853 | 0.844 | 0.879 | 0.838 |
| Ours | 0.892 | 0.876 | 0.921 | 0.872 |

From the overall results, different models show high classification performance in the task of user satisfaction prediction, but clear differences exist. GCN and GAT, as representative graph-based models, demonstrate stronger advantages compared with traditional sequence modeling methods. GAT outperforms GCN in ACC, F1-Score, and AUC. This indicates that the introduction of the attention mechanism helps capture heterogeneous relationships between interaction nodes and improves the ability to identify key features.

The performance of Transformer and 1DCNN shows the competitiveness of sequence modeling methods in this task. Transformer surpasses 1DCNN in all four metrics, with more obvious improvements in AUC and F1-Score. This reflects the advantage of multi-head self-attention in modeling long interaction sequences. However, due to the constraints of its input form, Transformer still has limitations in modeling complex interaction relationships. As a result, its overall performance does not exceed that of graph neural networks.

Although 1DCNN is slightly weaker than Transformer in terms of accuracy and robustness, in some metrics, it still approaches the level of GCN. This shows that convolutional structures have certain strengths in capturing local patterns and can effectively extract short-term interaction behaviors. However, when facing more complex and global dependencies, its expressive ability is clearly insufficient, which limits its potential in user satisfaction prediction tasks.

The proposed method achieves the best results in all evaluation metrics. ACC, F1-Score, AUC, and Precision are all significantly higher than those of the baseline models. This shows that structured modeling with graph neural networks, combined with further optimization mechanisms, can more comprehensively integrate user behaviors, interface elements, and contextual information. As a result, it provides more accurate support for satisfaction prediction. These results not only verify the adaptability and advantages of the method in complex interaction scenarios but also highlight its significance in improving the quality of user experience modeling.

This paper presents an experiment on the sensitivity of label noise rate to F1-Score, and the experimental results are shown in Figure 2.

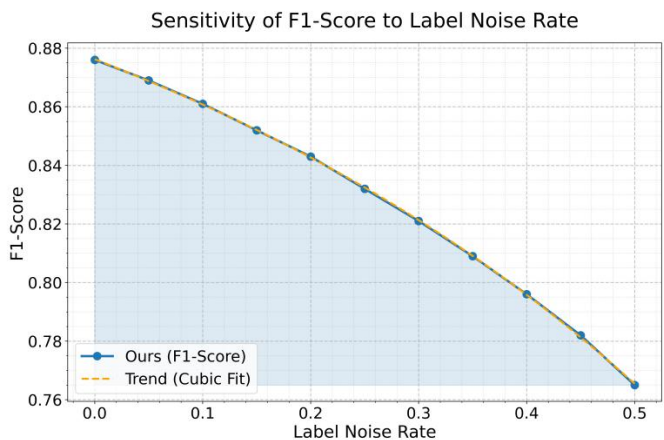

Figure 2. Sensitivity experiment of label noise rate to F1-Score

The experimental results show that as the label noise rate increases, the F1-Score exhibits a stable and continuous decline. This indicates that the user satisfaction classification model can maintain high accuracy in low-noise environments. However, as the proportion of incorrect labels increases, the overall performance of the model is significantly affected. This reflects the critical role of label quality in this task.

When the noise rate is low, the model can still capture the true patterns of user satisfaction through the propagation and aggregation of graph structural features. As a result, the decrease in F1-Score remains small, demonstrating a certain level of robustness. Once the noise rate exceeds 0.2, the decline accelerates, showing that incorrect labels interfere with the training signals. This makes it difficult for the model to learn the real relationships between users and interaction elements.

The fitted trend curve further shows that the performance drop is not linear but gradually accelerates. This means that in medium to high noise environments, the classification ability of the model deteriorates quickly. It highlights the importance of ensuring label reliability in satisfaction prediction tasks. With a high noise ratio, the model may overfit incorrect information and lose the ability to capture genuine interaction patterns.

Overall, this experiment reveals that data quality is crucial for graph neural network-based satisfaction prediction frameworks. Ensuring label accuracy is not only the basis for stable model performance but also affects the generalization ability of the model in complex human-computer interaction environments. The results emphasize the need for future studies to explore more robust modeling methods, such as noise-aware mechanisms or adversarial regularization, to reduce the negative impact of label uncertainty on user satisfaction modeling.

V. CONCLUSION

This study addresses the problem of user satisfaction classification and proposes a modeling framework based on graph neural networks. From the perspective of structured interaction relationships, it systematically characterizes the complex dependencies between users and systems. By constructing graph structures to represent user behaviors, interface elements, and their potential connections, the framework effectively overcomes the limitations of traditional methods that rely on linear or sequential modeling. It shows clear advantages in semantic representation and global dependency modeling, providing a more accurate and robust solution for automated satisfaction prediction. This approach not only enriches the theoretical system of interaction experience modeling but also offers a new research paradigm for related fields.

From both the experimental results and the characteristics of the method, the framework balances local detail and global semantics, and it handles the challenges of multi-source heterogeneous data and complex interaction environments. The overall improvement in performance metrics demonstrates the feasibility and effectiveness of applying graph neural networks to satisfaction prediction. By jointly modeling node features and edge relationships, the model shows strong classification ability and generalization capacity. This lays a solid foundation for further exploration of satisfaction modeling in more complex interaction scenarios. The structure-aware learning mechanism not only improves prediction accuracy but also enhances the interpretability of interaction patterns. At the application level, the significance of this study becomes evident. User satisfaction serves as a pivotal indicator for optimizing human-computer interaction, and accurate classification models can directly propel progress in various domains. For instance, in personalized recommendations, they can more effectively capture user interests and requirements, thereby enhancing recommendation outcomes. In intelligent customer service systems, they can promptly detect user experiences in real time and adjust response strategies adaptively. Similarly, in online education platforms and virtual reality environments, they can identify changes in learner or user satisfaction and facilitate improvements in interaction content and service design. Consequently, this study transcends mere methodology and assumes a crucial role in advancing intelligent interaction systems and optimizing user experiences.

Future research should extend and deepen this framework. One direction is to examine its applicability to larger-scale, multimodal, and cross-scenario data to ensure adaptability to expanding interaction environments. Another is to develop more robust mechanisms to address label noise and data uncertainty, ensuring stable and reliable classification results. In addition, as intelligent systems increasingly influence public services, healthcare, and social governance, the outcomes of satisfaction prediction will directly affect the sustainability and social value of human-computer interaction. Thus, this study provides not only a new methodology for the academic community but also technical support and practical guidance for future applications.